\documentclass[a4paper,10pt,unnumberedsections,twoside,]{LTJournalArticle}

\runninghead{}

\footertext{\textit{RISC-V Summit Europe, Paris, 12-15th May 2025}}

\usepackage{amsmath,amssymb,amsfonts}
\usepackage{algorithm2e}
\usepackage{graphicx}
\usepackage{textcomp}
\usepackage{xcolor}
\usepackage{enumitem}
\usepackage{tabularx}
\usepackage{booktabs}
\usepackage{multirow}
\usepackage{svg}
\usepackage{multicol}
\usepackage{adjustbox}
\usepackage{pdfpages}
\usepackage{subcaption}
\usepackage{biblatex}

\begin{document}

\title{Efficient Trace for RISC-V: Design, Evaluation, and Integration in CVA6}

\author{
    Umberto Laghi, Simone Manoni, Emanuele Parisi, Andrea Bartolini
}

\date{\footnotesize\textsuperscript{Department of Electrical, Electronic, and Information Engineering (DEI) -- University of Bologna, Italy}\vspace{-0.7mm}}

\renewcommand{\maketitlehookd}{

\begin{abstract}
\vspace{-1.9mm}
\noindent In this work, we present the design and evaluation of a Processor Tracing System compliant with the RISC-V Efficient Trace specification for Instruction Branch Tracing. We integrate our system into the host domain of a state-of-the-art edge architecture based on CVA6. The proposed Tracing System introduces a total overhead of 9.2\% in terms of resource utilization on a Xilinx VCU118 FPGA on the CVA6 subsystem while achieving an average compression rate of 95.1\% on platform-specific tests, compared to tracing each full opcode instruction.
\end{abstract}
}

\maketitle

\section{Introduction}
\vspace{-3.5mm}
In modern computing systems, understanding program execution is challenging, as software behavior and performance can deviate from expectations due to interactions with other cores, real-time events, suboptimal software implementations, or a combination of these factors.
Profiling techniques commonly used to analyze code and monitor program execution often involve significant trade-offs between being intrusive, enabling detailed debugging, or providing only high-level insight into execution performance~[1].

\textit{Instruction Branch Tracing} (IBT) allows continuous, non-intrusive, fine-grained monitoring of program execution by tracking program counter (PC) address deltas induced by \textit{special} instructions: jump, call, return, branch, interrupts, or exceptions.
RISC-V (RV) provides a standardized IBT mechanism known as Efficient Trace (E-Trace)~[2]. 
It splits the code into blocks, where each block is an instruction sequence bounded by two special instructions, whenever a discontinuity occurs, a dedicated hardware \textit{Trace Encoder} (TE) module generates a trace packet. 
The trace packet sequence is then processed by a software \textit{Trace Decoder} (TD) running on a host machine, which reconstructs the program execution by integrating the trace packet data with the program binary.

This work presents three main contributions: i) The design of a Tracing System (TS) compliant with the RISC-V E-Trace specification~[2]; ii) Its integration into a modern RISC-V edge platform based on the CVA6 core~[3]; iii) The evaluation of the proposed implementation in terms of achieved compression rate and FPGA resource utilization overhead. 

\vspace{-5mm}
\section{Architecture}
\vspace{-3.5mm}

\begin{figure}[ht]
    \centering
    \begin{subfigure}{0.45\textwidth}
        \centering
        \includegraphics[width=\linewidth]{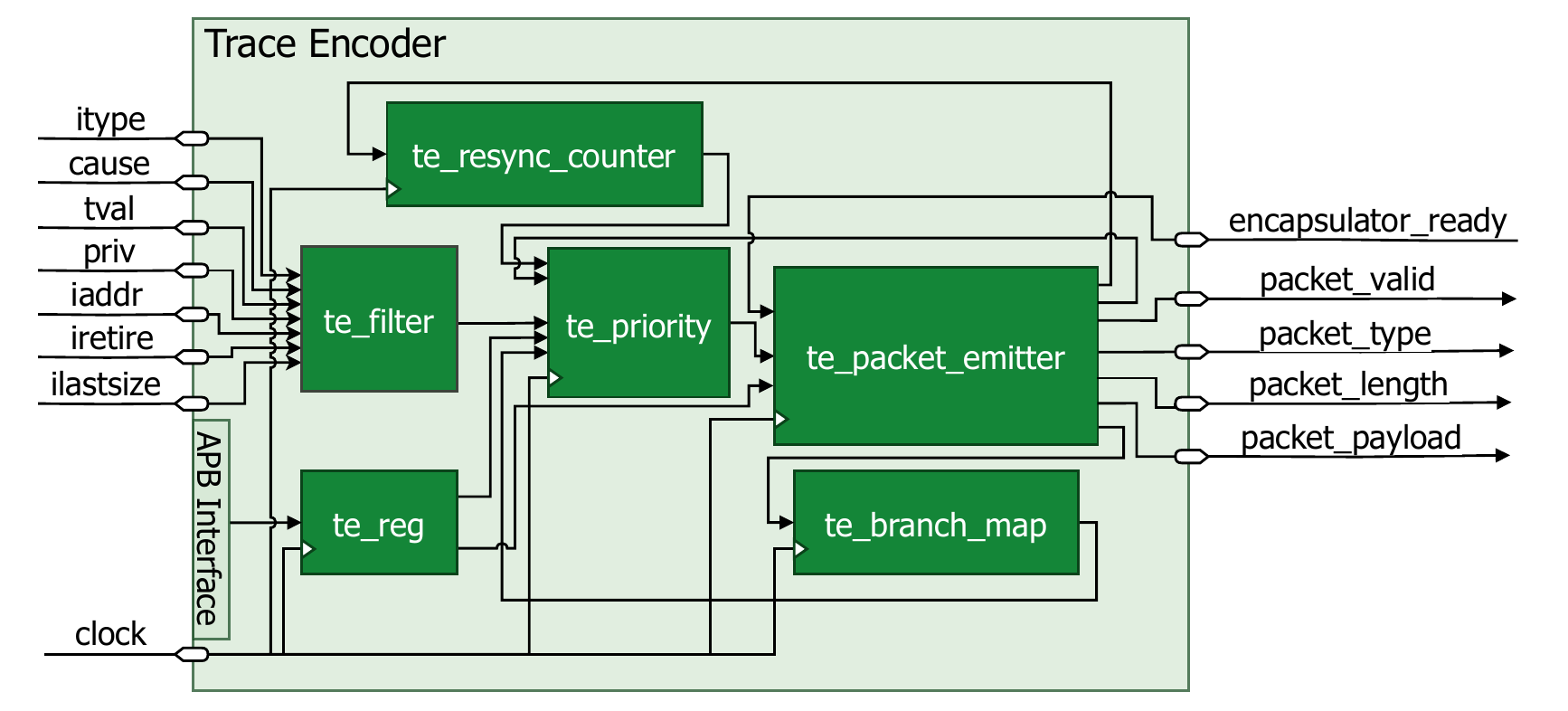}
        \caption{Trace Encoder architecture.}
        \label{fig:te}
    \end{subfigure}
    \begin{subfigure}{0.45\textwidth}
        \centering
        \includegraphics[width=\linewidth]{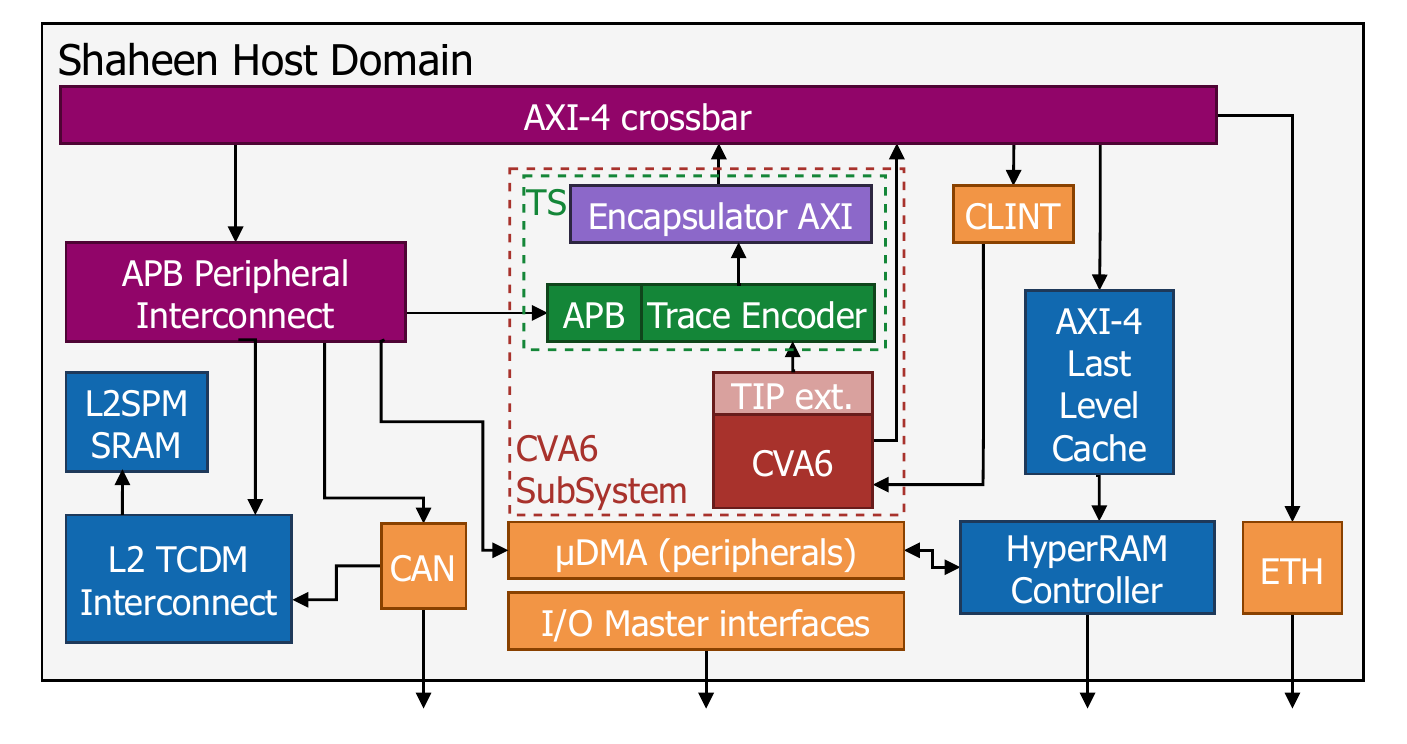}
        \caption{Shaheen host-domain extended with TS.}
        \label{fig:shaheen}
    \end{subfigure}
    \caption{TE architecture and integration inside Shaheen}
    \vspace{-5mm}
    \label{fig:arch}
\end{figure}
Figure~\ref{fig:te} illustrates the TE architecture. It takes as input the length of
the first instruction in the block (\texttt{iaddr}), the type
and length of the last instruction in the block (\texttt{itype},
\texttt{ilastsize}), the block length (\texttt{iretire}), the privilege
level (\texttt{priv}), and interrupt/exception details (\texttt{tval},
\texttt{cause}). The TE consists of three main building blocks, \texttt{te\_filter}, \texttt{te\_priority}, and \texttt{te\_packet\_emitter}, which are responsible for identifying trace blocks and generating E-trace packets. Additionally, it includes three support and configuration modules, namely \texttt{te\_reg}, \texttt{te\_resync\_counter}, and \texttt{te\_branch\_map}.
The \texttt{te\_filter} determines which information to trace. Its parameters are stored in the \texttt{te\_reg} module which manages user-configurable settings, including trace enablement, operating mode selection, and filter definitions to refine trace scope, and it can be configured via an APB interface.
The \texttt{te\_priority} module determines which packet to issue according to E-Trace.
The \texttt{te\_packet\_emitter} module is responsible for constructing the trace packets. 
It receives the packet type from the \texttt{te\_priority} module and collects the required data from the \texttt{te\_reg} module, as well as delayed TE inputs from registers.
The inputs are delayed by up to two cycles, allowing it to access them from current, previous and next cycle. 
Tracking discontinuities in these three states is essential to select the appropriate packet for output.
The \texttt{te\_branch\_map} module monitors branch instructions and keeps track of their results. 
It uses a 31-bit register to store branch results (the branch map) and a counter. 
When the branch map is full, the module requests the generation of a packet to report branch information, so no branch data is lost.
The \texttt{te\_resync\_counter} module tracks the number of packets emitted or clock cycles elapsed. 
When the predefined threshold is reached, it triggers a resynchronization request to the \texttt{te\_priority} module. 

A key feature of the TE is its ability to support processors that can retire multiple instructions per cycle. 
This is achieved by replicating block-specific inputs (\texttt{iaddr}, \texttt{itype}, \texttt{ilastsize}, \texttt{iretire}) and \texttt{te\_filter}, \texttt{te\_priority} and \texttt{te\_packet\_emitter} modules based on the number and type of discontinuities that can be retired by the CPU each cycle.

The E-Trace packets generated by the TE are transferred to an AXI4 encapsulator module via a valid/ready handshake. Upon receiving the payload, it generates AXI4 transactions to route trace packets through the system crossbar and transmit them over Ethernet.
Figure~\ref{fig:shaheen} shows the integration of our design into the host domain of Shaheen~[3], a state-of-the-art RISC-V platform featuring a CVA6 core for autonomous nano-drones.
We extended the CVA6 interface with a Trace Interface Port (TIP) that generates TE inputs according to E-Trace~[2].

\vspace{-5mm}
\section{Results and Conclusions}
\begin{figure}[ht]
\vspace{-2mm}
    \centering
    \includegraphics[width=0.6\columnwidth]{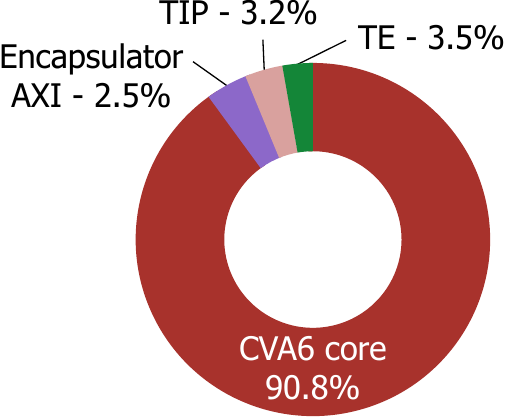}
    \caption{Resources utilization with respect to the CVA6 subsystem}
    \label{fig:area_utilization}
\end{figure}

\vspace{-3.5mm}
We conducted a preliminary evaluation of the TS by integrating it into Shaheen and implementing it on an FPGA emulator on the Xilinx VCU118 FPGA. We evaluated the area overhead and compression rate achieved using platform-specific benchmarks and regression tests. Figure~\ref{fig:area_utilization} shows the area overhead introduced by the Tracing System over CVA6. 
The TS consumes about 9.2\% of the total resources of the CVA6 subsystem (which includes both the CVA6 core and the TS) and about 10\% of the area compared to the CVA6 core alone. The TIP and Encapsulator AXI consume area primarily due to the registers that implement FIFOs, which are essential for elastic buffering. In area-constrained scenarios, the TS footprint can be reduced by instantiating smaller depth FIFOs. However, decreasing the FIFO depth may result in packet loss, which could affect the completeness of the captured trace.

In addition, the modules do not have an impact on the critical path, so there is no impact on operating frequency.

\begin{table}[t]
    \centering
    \begin{tabular}{lc}
        \toprule
        \multicolumn{1}{c}{Test name} & Compression rate \% \\ 
        \midrule
        
        \texttt{axi\_hyper\_fibonacci} & 99.8 \\ \hline
        \texttt{bypass\_cva6\_dco} & 99.5 \\ \hline
        \texttt{can} & 87.3 \\ \hline
        \texttt{dhrystone} & 98.4 \\ \hline
        \texttt{fp16\_matmul} & 99.7 \\ \hline
        \texttt{fp16-vec\_matmul} & 99.7 \\ \hline
        \texttt{hello} & 90.4 \\ \hline
        \texttt{kmeans} & 95.0 \\ \hline
        \texttt{l1\_test} & 99.7 \\ \hline
        \texttt{llc\_spm\_test} & 87.6 \\ \hline
        \texttt{mbox\_test} & 91.7 \\ \hline
        \texttt{mm} & 99.7 \\ \hline
        \texttt{sb\_macl\_444} & 99.7 \\ \hline
        \texttt{sb\_macl\_844} & 99.7 \\ \hline
        \texttt{timer} & 85.2 \\ \hline
        Average & 95.1 \\ 
        \bottomrule
    \end{tabular}
    \caption{Compression rate for each test} 
    \label{tab:compression_rate_tests}
    \vspace{-2.5mm}
\end{table}
Table \ref{tab:compression_rate_tests} presents the compression rate achieved by the tracing system, with an average compression rate of 95.1\% compared to tracing each instruction with its full-opcode allocation.

We developed a RV TS and integrated it into an edge platform based on CVA6 called Shaheen.
For future work, we plan to conduct more comprehensive benchmarks, enhance the TE with additional hardware features to achieve higher compression rates and advanced functional monitoring, and formally release the implementation open-source. Additionally, the development of the TD is currently ongoing.
\vspace{-2mm}

\scriptsize
\linespread{0.8}
\paragraph{Acknowledgements} This activity has been supported by the HE EU EUPEX (g.a. 101033975), DECICE (g.a. 101092582), and DARE (g.a. 101143421) projects, as well as the Italian Research Center on High Performance Computing, Big Data, and Quantum Computing.
\normalsize
\linespread{}

\vspace{-4mm}

\end{document}